\documentclass{article}
\usepackage{spconf}
\usepackage{graphicx}
\usepackage{svg}
\usepackage{bm}
\usepackage{cite}
\usepackage{amsmath,amssymb,amsfonts,mathrsfs}
\usepackage{bbm}
\usepackage{amsxtra}
\usepackage{threeparttable}
\usepackage{algorithm,algorithmic}
\usepackage{etoolbox,siunitx}
\usepackage{multirow, booktabs}
\usepackage{balance}
\usepackage{textcomp}
\usepackage{xcolor}
\usepackage{url}
\usepackage{pifont}
\usepackage{threeparttable}
\usepackage{hyperref}
\usepackage{makecell}
\usepackage[table]{xcolor} %
\robustify\bfseries

\newcommand{\mpsd}[2]{#1 {\scriptsize$\pm$ #2}}
\newcommand{\mb}[1]{\mathbf{#1}}

\newcommand{\1}[1]{\cellcolor{gray!40}#1} %
\newcommand{\2}[1]{\cellcolor{gray!20}#1} %

\title{SUNAC: Source-aware Unified Neural Audio Codec}
\name{{\shortstack[c]{
      Ryo Aihara$^{1,2}$,
      Yoshiki Masuyama$^{1}$,
      Francesco Paissan$^{1,3}$,\\
      Fran\c{c}ois G.\ Germain$^{1}$,
      Gordon Wichern$^{1}$,
      Jonathan Le Roux$^{1}$%
      \thanks{This work was done while F. Paissan was an intern at MERL.}
}}}
\address{
$^{1}$Mitsubishi Electric Research Laboratories (MERL), Cambridge, USA \\
$^{2}$Information Technology R\&D Center, Mitsubishi Electric Corporation, Kanagawa, Japan, \\
$^{3}$University of Trento, Trento, Italy
}
\begin{document}
\ninept
\setlength{\abovedisplayskip}{6pt}
\setlength{\belowdisplayskip}{6pt}
\maketitle
\begin{abstract}
Neural audio codecs (NACs) provide compact representations that can be leveraged in many downstream applications, in particular large language models. Yet most NACs encode mixtures of multiple sources in an entangled manner, which may impede efficient downstream processing in applications that need access to only a subset of the sources (e.g., analysis of a particular type of sound, transcription of a given speaker, etc). To address this, we propose a source-aware codec that encodes individual sources directly from mixtures, conditioned on source type prompts. This enables user-driven selection of which source(s) to encode, including separately encoding multiple sources of the same type (e.g., multiple speech signals). Experiments show that our model achieves competitive resynthesis and separation quality relative to a cascade of source separation followed by a conventional NAC, with lower computational cost.
\end{abstract}

\section{Introduction}
\label{sec:intro}
\vspace{-1.0mm}

With the advent of large language models, end-to-end discrete neural audio codecs (NACs) 
have been widely investigated as a front end for converting audio signals 
into discrete text-like tokens~\cite{mousavi2025discrete,mousavi2024dasb}. 
A typical setup employs generative adversarial network (GAN)-based training, a convolutional encoder for waveform analysis, 
a residual vector quantization (RVQ) module for discretization, 
and a (transposed-)convolutional decoder for 
waveform synthesis~\cite{zeghidour2021soundstream,defossez2022highfi,kumar2023highfidelity,zhang2024speechtokenizer}. 

Because speech communication over a channel is a primary application of NACs, 
their noise robustness has received considerable attention~\cite{casebeer2021enhancing, yang2021source, chae2025towards, luo2025decodec}. 
In real-world scenarios, however, received audio often contains multiple concurrent sources, such as speech, music, and environmental sounds. %
Most conventional NACs are trained without source awareness, 
and thus encode mixtures without disentangling the constituent sources. 
We posit that encoding and transmitting mixtures in this manner is suboptimal for downstream tasks 
that predominantly target a single source (e.g., meeting summarization~\cite{shang2024endtoendspeechsummarization, matsuura2024sentencewises}, full-duplex voice assistants~\cite{kyutai2024moshi,ohashi2025jmoshi,hu2025salmduplex}, acoustic event detection~\cite{wang2024leveraginglanguagemodel,yin2025exploring}, music-language models~\cite{zhang2025instructmusicgen, wang2025usam}, etc).

\begin{figure}[t]
  \centering
  \includegraphics[width=0.48\textwidth]{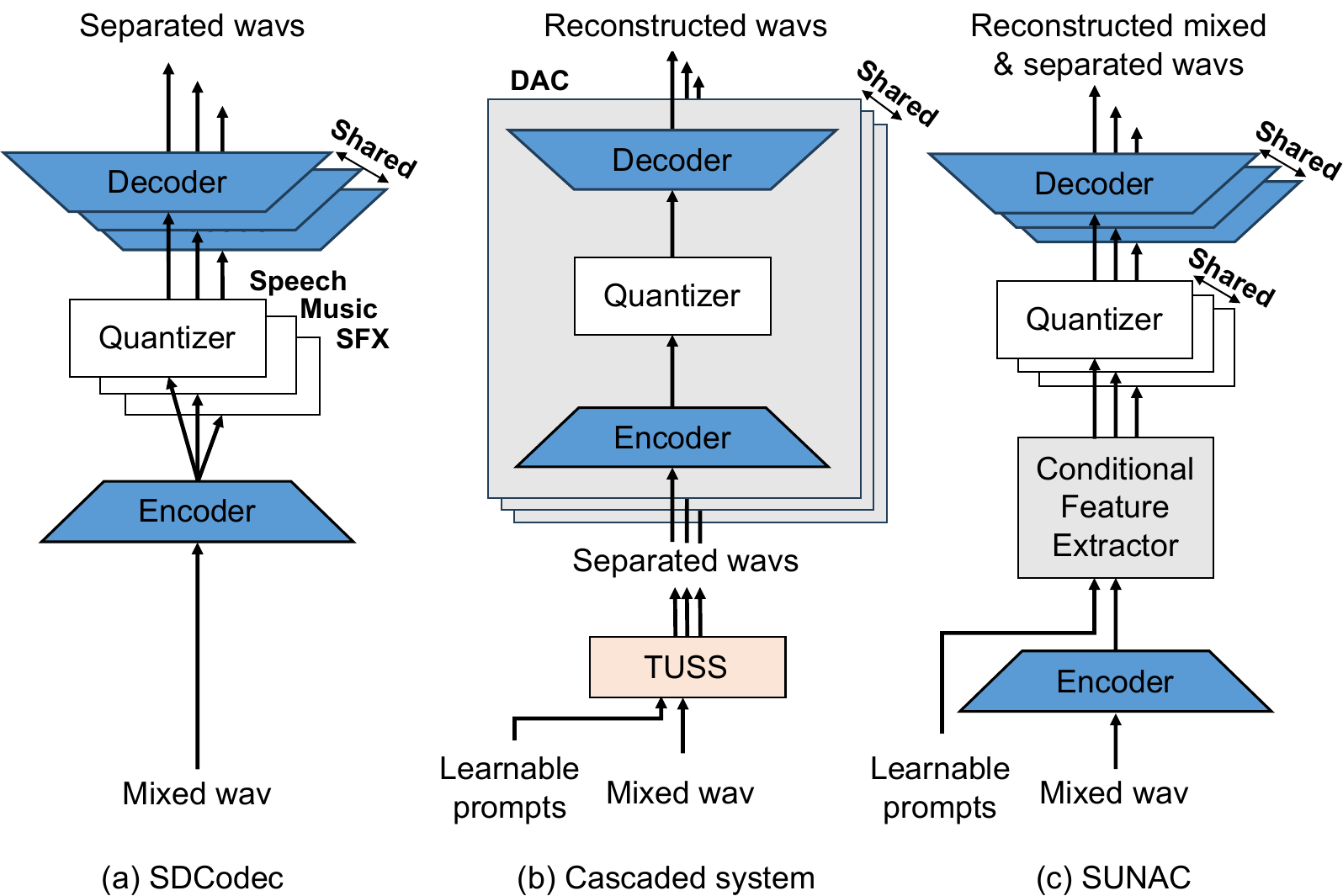}
  \vspace{-6mm}
  \caption{Overview of (a) SDCodec~\cite{bie2025learning}, (b) the proposed cascaded system, and (c) the proposed SUNAC model.}
  \label{fig:overview}
  \vspace{-6mm}
\end{figure}

SDCodec~\cite{bie2025learning} addressed this challenge by augmenting conventional NACs 
with parallel, source-aware RVQ modules (Fig.~\ref{fig:overview}(a)). 
This straightforward design can separately encode and reconstruct sources 
from mixtures of speech, music, and sound effects (SFX). 
However, because the separation capacity is tied to distinct source-aware RVQs, 
the method cannot separate mixtures of sources from the same category (e.g., two concurrent speakers).

To separately encode sources (including those from the same category), we first propose a cascaded architecture that couples a unified source separator~\cite{pons2024gass, manilow2020hierarchical, petermann2023hyperbolic}
with a conventional NAC, i.e., first separate, then encode (Fig.~\ref{fig:overview}(b)).  For the separator, we use the recently introduced task-aware unified source separator (TUSS)~\cite{saijo2025task-aware}, 
which consolidates multiple separation tasks within a single model and 
selects the desired operation via lightweight prompting.
However, this cascaded design is not computationally efficient, since both the separator and the NAC independently derive compact representations from the same input audio, leading to redundant computation.

We thus also propose a source-aware unified neural audio codec (SUNAC) (Fig.~\ref{fig:overview}(c)), where 
all components are trained end-to-end with joint optimization, in contrast to the cascaded system.
SUNAC performs prompt-based source feature extraction directly in the latent space, 
after which a quantizer estimates codes on the separated features. 
To address permutation ambiguity when processing multiple sources of the same type,
we apply permutation invariant training (PIT)~\cite{Hershey2016deep, Kolbaek2017multitalker} in the feature space. 
Both the cascaded model and SUNAC can separate multiple sources from the same or different types. 
Moreover, the prompting mechanism removes any predefined cap on the number of sources, 
allowing the model to scale to an arbitrary number in principle.
Experimental results show that SUNAC is comparable to SDCodec in scenarios without multiple sources from the same category, 
that it can also encode and reconstruct each speech source in multi-speaker mixtures. %
Furthermore, SUNAC achieves performance on par with the cascaded architecture 
while offering lower computational cost.

\vspace{-0.25cm}
\section{Source-aware Unified Neural Audio Codec}
\vspace{-0.1cm}
\subsection{Problem Setup}
\label{sec:setup}
\vspace{-0.05cm}
We consider an input waveform $\mb{x}=\sum_n \mb{s}^{(p_n)}_n \in \mathbb{R}^{L}$ consisting of $N\geq 1$ sources $\mb{s}^{(p_n)}_n$, each associated with a source type described by a prompt $p_n\in \mathcal{T}$. In this paper, we consider sources to be either speech, music, SFX, or a mixture of these, such that $\mathcal{T}=\{\texttt{<Speech>}, \texttt{<Music>}, \texttt{<SFX>}, \texttt{<Mix>}\}$. Multiple sources may be of the same type. Our goal is to extract codes for one or more of these sources, specified by a set of desired prompts.

\vspace{-0.1cm}
\subsection{SDCodec}
\label{sec:related}
\vspace{-0.05cm}
SDCodec~\cite{bie2025learning}, illustrated in Fig.~\ref{fig:overview}(a), considers a restricted version of our setup where $N=3$ and there is one source each of speech, music, and SFX.
It extends the Descript audio codec (DAC)~\cite{kumar2023highfidelity} to handle parallel processing of mixture components %
via the insertion of source-aware RVQ modules after the convolutional encoder. 
This design compels all sources to share a common encoder-derived feature space, 
while enabling per-source quantization pathways; consequently, 
orthogonality across source features is encouraged, 
similarly to the approach in~\cite{yang2021source}.
By contrast, our proposed SUNAC estimates a prompt-conditioned feature space, 
removing the need to enforce such orthogonality. 
Moreover, SDCodec reconstructs mixtures by decoding the sum of the per-source quantized features, imposing additivity in the quantized space, whereas SUNAC reconstructs mixtures directly by prompting the model with a \texttt{<Mix>} prompt. Finally, SDCodec cannot separate multiple sources of the same type.

\vspace{-0.1cm}
\subsection{Separation and NAC Cascade}
\label{sec:cascade}
\vspace{-0.05cm}
As a straightforward approach for our problem setup, we propose to consider a cascaded system which first separates the mixture into the desired source-specific waveform signals, before encoding them. A natural choice is to employ TUSS~\cite{saijo2025task-aware} as a front end to DAC, as illustrated in Fig.~\ref{fig:overview}(b).
In TUSS, the mixture waveform is encoded via short-time Fourier transform (STFT) and 
a band-split encoder. The encoded features and learnable prompts indicating the target source type are transformed by TF-Locoformer layers~\cite{saijo2024tf-locoformer}. 
The transformed features are conditioned by element-wise multiplication 
with the transformed prompts, after appropriate broadcast. 
The conditioned features are then refined by more TF-Locoformer layers and 
mapped back to the time domain using 
an inverse band-split operation and inverse STFT. By default, the TUSS-DAC cascade is quite inefficient, but we explore recent advances in computational efficiency for both the separation~\cite{paissan2025fastuss} and NAC~\cite{kyutai2024moshi} components.

\begin{figure}[t]
  \centering
  \includegraphics[width=0.4\textwidth]{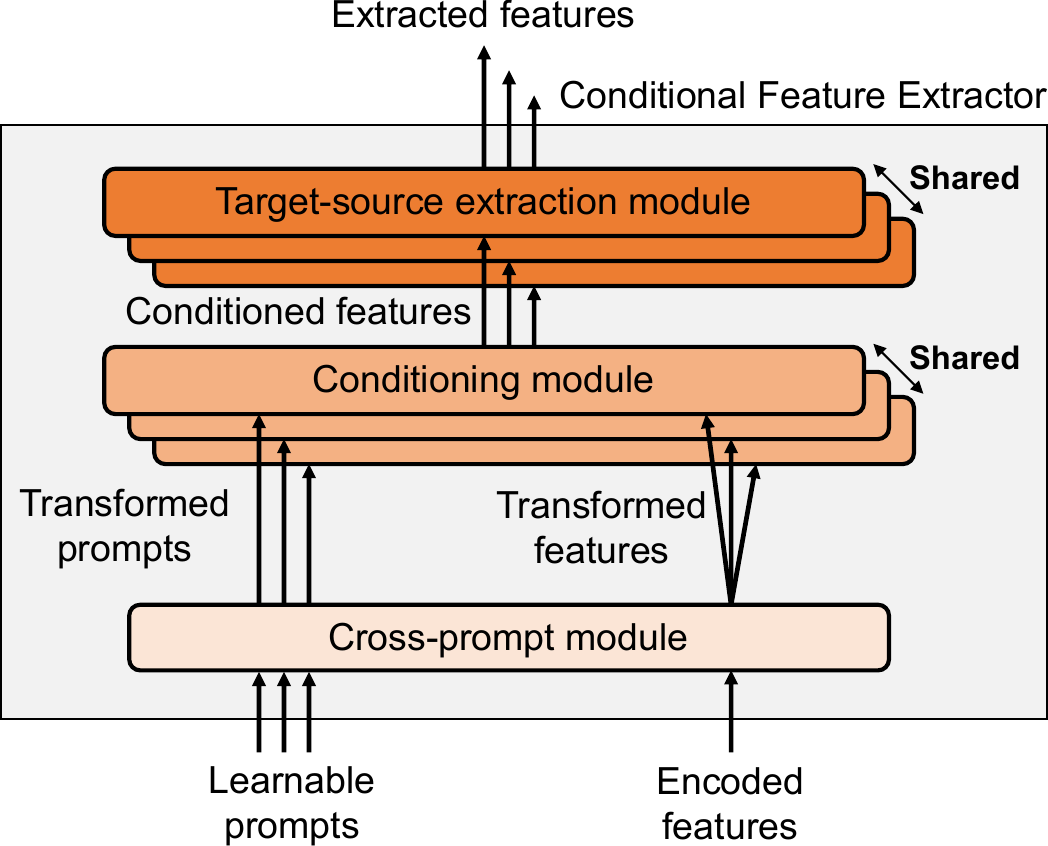}
  \vspace{-3mm}
  \caption{Detailed architecture of conditional feature extractor}
  \label{fig:separator}
  \vspace{-4mm}
\end{figure}

\vspace{-0.1cm}
\subsection{SUNAC}
\label{sec:method}
\vspace{-0.05cm}
To avoid redundant processing, we propose to replace the explicit separation in the cascaded system by a conditional feature extractor in the feature space, leading to the integrated SUNAC architecture illustrated in Fig.~\ref{fig:overview}(c). 
The encoder maps the input waveform $\mb{x}$  
into a continuous time–frequency (TF)-like representation $\mb{X} \in \mathbb{R}^{F \times T}$, where $F$ and $T$ denote the feature dimension and the number of frames; we adopt a convolutional design following prior work~\cite{kumar2023highfidelity, bie2025learning}.
The conditional feature extractor estimates separated TF-representation based on input learnable prompts,
as explained in detail below.
The quantizer and decoder are source agnostic, i.e., shared across all sources.
The quantizer uses a multi-layer RVQ with projection~\cite{kumar2023highfidelity} 
to discretize the separated TF representations. 
The decoder then takes the quantized TF features and estimates waveforms $\mathbf{\hat{s}} \in \mathbb{R}^{N \times L}$, 
where $N$ is the number of prompts provided to the conditional feature extractor. 
Our decoder combines Transformer~\cite{vaswani2023attention} and convolutional layers, whereas the SDCodec decoder is convolution-only, thus incurring higher computational cost.

\begin{table}[t]
\vspace{-6pt}
\centering
    \sisetup{
        detect-weight,
        mode=text,
        tight-spacing=true,
        round-mode=places,
        round-precision=1,
        table-format=2.2,
        table-number-alignment=center
    }
    \caption{Comparison of model parameters and computational cost (MAC) measured per 1.0 second in GMACs. 
Const denotes the part of the MAC value independent of the number of Sources, while Per source represents the part that scales linearly with the number of Sources. 
The total MAC for $N$ sources is given by $\text{Const} + (\text{Per source}) \times N$, where $N$ is the number of sources.}
    \label{table:params_macs}
    \vspace{1pt}
    \begin{tabular}{l S S S}
        \toprule
        Method & {Params (M)} & {Const [G]} & {Per source [G]} \\
        \midrule
        TUSS                      & 11.10 & 21.10     & 10.50 \\
        FasTUSS                   & 11.10 & 4.13      & 2.06  \\
        \cmidrule(lr){1-4}
        DAC                       & 74.10 & {}        & 41.00 \\
        DACT                      & 66.42 & {}        & 12.88 \\
        \cmidrule(lr){1-4}
        TUSS-DAC                  & 85.20 & 21.10     & 51.50 \\
        FasTUSS-DACT              & 77.52 & 4.13      & 14.94 \\
        \cmidrule(lr){1-4}
        SDCodec                   & 74.82 & 12.56     & 28.44 \\
        SDCodecT                  & 67.06 & \2{3.91}  & \1{8.95}  \\
        SUNAC                     & 69.17 & \1{3.50}  & \2{9.45}  \\
        \bottomrule
    \end{tabular}
    \vspace{-5mm}
\end{table}

\begin{table*}[t]
    \vspace{-6pt}
    \caption{Reconstruction results for isolated signals of each source type.}
    \label{tab:reconst_nonmix}
    \centering
    \sisetup{
        detect-weight,
        mode=text,
        round-mode=places,
        round-precision=2,
        tight-spacing=true,
        table-number-alignment=center,
        separate-uncertainty = true
    }
    \vspace{1pt}
    \begin{tabular}{lccc ccc ccc}
        \toprule
        & \multicolumn{2}{c}{\textbf{Speech}} & \multicolumn{2}{c}{\textbf{Music}} & \multicolumn{2}{c}{\textbf{SFX}} \\
        \cmidrule(lr){2-3}\cmidrule(lr){4-5}\cmidrule(lr){6-7}
        \textbf{Model} &
        {\textbf{SI-SDR $\uparrow$}} & {\textbf{VisQOL $\uparrow$}} &
        {\textbf{SI-SDR $\uparrow$}} & {\textbf{VisQOL $\uparrow$}} &
        {\textbf{SI-SDR $\uparrow$}} & {\textbf{VisQOL $\uparrow$}} \\
        \midrule
        DAC                   & \2{\mpsd{8.85}{3.33}} & \mpsd{4.64}{0.16}     & \mpsd{7.61}{3.44}     & \1{\mpsd{4.63}{0.04}} & \mpsd{2.94}{5.60}     & \1{\mpsd{4.60}{0.05}} \\
        DACT                  & \mpsd{8.42}{3.90}     & \1{\mpsd{4.79}{0.07}} & \1{\mpsd{8.66}{4.23}} & \2{\mpsd{4.57}{0.06}} & \2{\mpsd{3.16}{5.77}} & \mpsd{4.54}{0.07}     \\
        SDCodec$^\dagger$     & \mpsd{7.24}{4.45}     & \mpsd{4.63}{0.16}     & \mpsd{7.65}{4.09}     & \mpsd{4.56}{0.07}     & \mpsd{2.24}{6.24}     & \mpsd{4.54}{0.09}     \\
        SDCodec               & \1{\mpsd{9.11}{3.87}} & \mpsd{4.67}{0.13}     & \mpsd{6.63}{4.36}     & \mpsd{4.23}{0.22}     & \1{\mpsd{3.36}{5.17}} & \2{\mpsd{4.55}{0.10}} \\
        SDCodecT              & \mpsd{8.20}{4.29}     & \mpsd{4.68}{0.12}     & \mpsd{7.45}{4.57}     & \mpsd{4.23}{0.22}     & \mpsd{2.81}{5.67}     & \mpsd{4.54}{0.07}     \\
        SUNAC                 & \mpsd{7.99}{4.39}     & \2{\mpsd{4.73}{0.11}} & \2{\mpsd{7.76}{4.11}} & \mpsd{4.56}{0.06}     & \mpsd{2.17}{6.63}     & \mpsd{4.54}{0.07}     \\
        \bottomrule
    \end{tabular}
    \vspace{-6pt}
\end{table*}

\begin{table*}[ht]
    \vspace{-8pt}
    \centering
    \sisetup{
        detect-weight,
        mode=text,
        round-mode=places,
        round-precision=2,
        tight-spacing=true,
        table-number-alignment=center,
        separate-uncertainty = true
    }
    \caption{Reconstruction results for the mixture and separated sources from \{\texttt{<Speech>}, \texttt{<Music>}, \texttt{<SFX>}\} (no repeated prompt).}
    \label{tab:seprate_w/o_perm}
    \vspace{1pt}
    \begin{tabular}{lcccccccc}
        \toprule
         & \multicolumn{2}{c}{\textbf{Mix}} & \multicolumn{2}{c}{\textbf{Speech}} & \multicolumn{2}{c}{\textbf{Music}} & \multicolumn{2}{c}{\textbf{SFX}} \\
        \cmidrule(lr){2-3}\cmidrule(lr){4-5}\cmidrule(lr){6-7}\cmidrule(lr){8-9}
        \textbf{Model} &
        \textbf{SI-SDR $\uparrow$} & \textbf{VisQOL $\uparrow$} &
        \textbf{SI-SDR $\uparrow$} & \textbf{VisQOL $\uparrow$} &
        \textbf{SI-SDR $\uparrow$} & \textbf{VisQOL $\uparrow$} &
        \textbf{SI-SDR $\uparrow$} & \textbf{VisQOL $\uparrow$} \\
        \midrule
        TUSS\text{-}DAC               & --                    & --                    & \mpsd{13.07}{2.46}     & \mpsd{3.72}{0.38}     & \mpsd{4.70}{4.18}     & \mpsd{4.25}{0.18}     & \mpsd{4.82}{5.17}     & \mpsd{4.19}{0.18} \\
        \midrule
        FasTUSS\text{-}DACT           & --                    & --                    & \1{\mpsd{12.29}{2.50}} & \mpsd{3.53}{0.39}     & \1{\mpsd{3.92}{4.18}} & \2{\mpsd{4.26}{0.16}} & \1{\mpsd{3.80}{5.38}} & \mpsd{4.17}{0.17}     \\
        SDCodec$^\dagger$             & \mpsd{6.39}{3.19}     & \mpsd{4.52}{0.05}     & \mpsd{10.78}{2.99}     & \mpsd{3.50}{0.43}     & \mpsd{1.74}{3.57}     & \1{\mpsd{4.26}{0.13}} & \2{\mpsd{2.17}{4.33}} & \1{\mpsd{4.20}{0.15}} \\
        SDCodec                       & \1{\mpsd{8.24}{2.83}} & \1{\mpsd{4.56}{0.05}} & \mpsd{11.40}{3.08}     & \mpsd{3.64}{0.41}     & \mpsd{1.21}{3.58}     & \mpsd{4.10}{0.21}     & \mpsd{1.26}{4.27}     & \2{\mpsd{4.18}{0.17}} \\
        SDCodecT                      & \2{\mpsd{7.30}{3.06}} & \2{\mpsd{4.54}{0.06}} & \mpsd{11.32}{2.89}     & \2{\mpsd{3.64}{0.37}} & \mpsd{1.75}{4.45}     & \mpsd{4.09}{0.22}     & \mpsd{1.42}{4.76}     & \mpsd{4.09}{0.22}     \\
        SUNAC                         & \mpsd{6.48}{3.11}     & \mpsd{4.52}{0.06}     & \2{\mpsd{11.56}{3.00}} & \1{\mpsd{3.68}{0.36}} & \2{\mpsd{1.98}{4.62}} & \mpsd{4.14}{0.20}     & \mpsd{2.10}{4.94}     & \mpsd{4.16}{0.19}     \\
        \bottomrule
    \end{tabular}
    \vspace{-3mm}
\end{table*}
As shown in Fig.~\ref{fig:separator}, the conditional feature extractor consists of a cross-prompt module, a conditioning module, and a target-source extraction module. 
The cross-prompt module takes as input the encoded TF representation $\mb{X} \in \mathbb{R}^{F \times T}$ and 
learnable prompt vectors $\mb{P} \in \mathbb{R}^{F \times N}$ corresponding to the $N$ prompts $(p_n)_n$.
To enable reconstruction of multi-source mixtures, 
the model is trained to reconstruct the mixture as is when \texttt{<Mix>} is supplied.
We concatenate the $N$ prompts to $\mathbf{X}$ along the time axis, apply Transformer layers along time, 
and then split off the first $N$ tokens to yield transformed prompts $\mb{P}' \in \mathbb{R}^{F \times N}$ and
transformed features $\mb{X}' \in \mathbb{R}^{F \times T}$.
With positional encoding~\cite{su2023roformer} and self-attention, prompts with the same content but different positions produce different representations.
As a result, the input TF features are influenced by the prompts and mapped 
into a space that facilitates conditional extraction.
The conditioning module applies feature-wise linear modulation (FiLM)~\cite{perez2018film} 
with residual connection to the transformed features $\mathbf{X}'$ using each transformed prompt $\mathbf{P}'_{n} \in \mathbb{R}^{F}$. Using arbitrary trainable functions $f$ and $h$ here implemented as simple linear transformations shared across all prompts, the FiLM output is computed as:
\begin{align}
    \mathrm{FiLM}(\mb{X}'|\mb{P}'_n) &= f(\mb{P}'_n) \odot \mb{X}' + h(\mb{P}'_n),\label{eq:film}
\end{align}
where %
$\odot$
denotes the element-wise product. 
We also evaluated TUSS-style conditioning via element-wise multiplication with the prompt broadcast over time~\cite{saijo2025task-aware}, but that proved less effective in our experiments.
Finally, we implement the target-source extraction module with Transformer layers that refine the conditioned TF representation, 
shared across all prompts.

\subsection{Training objective}
\label{subsec:objective}
We can train SUNAC with a permutation-invariant objective~\cite{Hershey2016deep, Kolbaek2017multitalker}:
\begin{align}
\mathcal{L}_{\text{SUNAC}} = \min_{\pi \in \tilde{\mathcal{P}}_S} \sum_{i=1}^{S} \mathcal{L}_{\text{DAC}} \left( s_i, \hat{s}_{\pi(i)} \right) + \mathcal{L}_{\text{DAC}} \left( s_{\text{mix}}, \hat{s}_{\text{mix}} \right),
\end{align}
where $s_i$ and $\hat{s}_{\pi(i)}$ are the $i$-th ground-truth source and its assigned estimate, 
while $s_{\text{mix}}$ and $\hat{s}_{\text{mix}}$ are the ground-truth and estimated mixtures, $S$ is the number of sources, and $\tilde{\mathcal{P}}_S$ is the subset of permutations of $\{1,\dots,S\}$ that only permute indices corresponding to prompts of the same type, leaving others fixed; $\pi \in \tilde{\mathcal{P}}_S$ can thus align, e.g., multiple \texttt{<Speech>} estimates to their appropriate references.
$\mathcal{L}_{\text{DAC}}$ denotes the DAC loss, which consists of a weighted sum of 
multi-scale mel-spectrogram loss, adversarial loss, codebook loss, commitment loss, and discriminator loss.
The discriminator consists of a multi-period discriminator~\cite{kong2020hifigan}
and a complex multi-scale STFT discriminator~\cite{jang2021univnet}.
The entire model is trained in a generative adversarial manner, as outlined in~\cite{kumar2023highfidelity}.

{\allowdisplaybreaks
In practice, evaluating all components of the DAC loss over all permutations is computationally prohibitive. 
Instead, we determine the permutation using a scale-invariant signal-to-distortion ratio (SI-SDR)–based criterion~\cite{leroux2019sdr} and compute the DAC loss only for the output–reference assignment that minimizes this SI-SDR criterion as follows:
\begin{align}
\pi^\star &= \operatorname*{arg\,max}_{\pi \in \tilde{\mathcal{P}}_S} \sum_{i=1}^{S} \text{SI-SDR}\!\left( s_i, \hat{s}_{\pi(i)} \right) \\
\mathcal{L}_{\text{SUNAC}} &= \sum_{i=1}^{S} \mathcal{L}_{\text{DAC}} \left( s_i, \hat{s}_{{\pi^\star}(i)} \right) + \mathcal{L}_{\text{DAC}} \left( s_{\text{mix}}, \hat{s}_{\text{mix}} \right)
\end{align}
where $\pi^\star$ denotes the optimal permutation of the estimated signals with respect to the reference sources, restricted to permutations among sources of the same type (cross-type indices fixed). %
}

\begin{table}[b]
    \vspace{-18pt}
    \centering
    \sisetup{
        detect-weight,
        mode=text,
        round-mode=places,
        round-precision=2,
        tight-spacing=true,
        table-number-alignment=center,
        separate-uncertainty = true
    }
    \caption{Separated results from \{\texttt{<Speech>},\texttt{<Speech>}\}.}
    \label{tab:speech_results}
    \vspace{1pt}
    \begin{tabular}{lcc}
        \toprule
        \textbf{Model} &
        {\textbf{SI-SDR $\uparrow$}} &
        {\textbf{VisQOL $\uparrow$}} \\
        \midrule
        TUSS\text{-}DAC               & \mpsd{13.35}{3.80}     & \mpsd{4.08}{0.39} \\
        \midrule
        FasTUSS\text{-}DACT           & \2{\mpsd{10.73}{4.66}} & \2{\mpsd{3.83}{0.46}} \\
        SDCodec$^\dagger$             & \mpsd{\phantom{1}0.00}{2.83}      & \mpsd{3.04}{0.61}     \\
        SDCodec                       & \mpsd{\phantom{1}0.00}{2.83}      & \mpsd{3.04}{0.62}     \\
        SDCodecT                      & \mpsd{\phantom{1}0.00}{2.83}      & \mpsd{3.09}{0.59}     \\
        SUNAC                         & \1{\mpsd{11.80}{3.07}} & \1{\mpsd{4.12}{0.42}} \\
        \bottomrule
    \end{tabular}
\end{table}

\begin{table*}[t]
    \vspace{-6pt}
    \centering
    \sisetup{
        detect-weight,
        mode=text,
        round-mode=places,
        round-precision=2,
        tight-spacing=true,
        table-number-alignment=center,
        separate-uncertainty = true
    }
    \caption{Reconstruction results for mixed source and separated results from \{\texttt{<Speech>}, \texttt{<Speech>}, \texttt{<Music>}, \texttt{<SFX>}\}.}
    \label{tab:two_speech_music_sfx}
    \begin{tabular}{lcccccccc}
        \toprule
         & \multicolumn{2}{c}{\textbf{Mix}} & \multicolumn{2}{c}{\textbf{Speech}} & \multicolumn{2}{c}{\textbf{Music}} & \multicolumn{2}{c}{\textbf{SFX}} \\
        \cmidrule(lr){2-3}\cmidrule(lr){4-5}\cmidrule(lr){6-7}\cmidrule(lr){8-9}
        \textbf{Model} &
        \textbf{SI-SDR $\uparrow$} & \textbf{VisQOL $\uparrow$} &
        \textbf{SI-SDR $\uparrow$} & \textbf{VisQOL $\uparrow$} &
        \textbf{SI-SDR $\uparrow$} & \textbf{VisQOL $\uparrow$} &
        \textbf{SI-SDR $\uparrow$} & \textbf{VisQOL $\uparrow$} \\
        \midrule
        TUSS\text{-}DAC               & --                     & --                     & \mpsd{9.07}{3.38}      & \mpsd{3.40}{0.47}     & \mpsd{2.75}{3.96}      & \mpsd{4.20}{0.17}     & \mpsd{3.05}{5.23}      & \mpsd{4.13}{0.18}     \\
        \midrule
        FasTUSS\text{-}DACT            & --                     & --                     & \2{\mpsd{6.98}{3.92}}  & \2{\mpsd{3.08}{0.38}} & \1{\mpsd{2.09}{3.82}}  & \1{\mpsd{4.21}{0.16}} & \1{\mpsd{2.07}{5.37}}  & \mpsd{4.11}{0.18}     \\
        SDCodec$^\dagger$             & \mpsd{6.55}{2.59}      & \mpsd{4.49}{0.06}      & \mpsd{-0.95}{3.29}     & \mpsd{2.58}{0.53}     & \mpsd{-0.69}{3.64}     & \2{\mpsd{4.20}{0.13}} & \mpsd{-0.15}{4.69}     & \1{\mpsd{4.15}{0.15}} \\
        SDCodec                       & \1{\mpsd{8.39}{2.31}}  & \1{\mpsd{4.54}{0.05}}  & \mpsd{-1.00}{3.34}     & \mpsd{2.64}{0.54}     & \mpsd{-1.62}{3.77}     & \mpsd{4.07}{0.21}     & \mpsd{-0.96}{4.44}     & \2{\mpsd{4.12}{0.17}} \\
        SDCodecT                      & \2{\mpsd{7.45}{2.52}}  & \2{\mpsd{4.51}{0.06}}  & \mpsd{-0.95}{3.58}     & \mpsd{2.60}{0.56}     & \mpsd{-1.15}{4.45}     & \mpsd{4.07}{0.22}     & \mpsd{-0.61}{4.81}     & \mpsd{4.11}{0.17}     \\
        SUNAC                         & \mpsd{6.38}{2.54}      & \2{\mpsd{4.51}{0.06}}  & \1{\mpsd{7.46}{3.41}}  & \1{\mpsd{3.33}{0.45}} & \2{\mpsd{0.15}{4.29}}  & \mpsd{4.11}{0.20}     & \2{\mpsd{0.25}{4.97}}  & \mpsd{4.11}{0.19}     \\
        \bottomrule
    \end{tabular}
    \label{tab:mix_speech_music_sfx_v2}
    \vspace{-3mm}
\end{table*}

\section{Experiments}
\label{sec:experiment}
\vspace{-.1cm}
\subsection{Experimental conditions}
\label{subsec:conditions}
We compare %
the following methods:

\noindent   \textbf{DAC}~\cite{kumar2023highfidelity}: 
        Single-source reconstruction baseline.
        The encoder consists of strided downsampling convolutional blocks with factors $(2,4,5,8)$, resulting in a token rate of $50\,\text{Hz}$ for $16\,\text{kHz}$ audio.
        Following the encoder, a twelve-layer RVQ module is applied, with each layer using a 1024-entry codebook.
        The decoder consists of upsampling convolutional blocks that reconstruct the waveform.
        
\noindent    \textbf{DACT}:
        Updated single-source reconstruction model. 
        Relative to DAC, the encoder’s convolutional base latent dimension is reduced from 64 to 32, and three Transformer layers (1024 hidden units, 8 attention heads) are added after the convolutional encoder. 
        In the decoder, the base latent dimension is reduced from 1536 to 768, and three Transformer layers (1024 hidden units, 8 heads) are inserted before the convolutional decoder. 
        The resulting architecture is a non-causal version of Mimi~\cite{kyutai2024moshi} that omits semantic tokens.
\noindent    \textbf{SDCodec}~\cite{bie2025learning}: 
        Extension of DAC separately handling three source types
        (see Section~\ref{sec:related}). Its encoder and decoder share the same architecture as the 16 kHz version of DAC. Unlike DAC, SDCodec uses three domain-specific RVQs, corresponding to speech, music, and sfx. 
        Each RVQ module is identical to that in DAC.

\noindent    \textbf{SDCodecT}:
        Updated variant of SDCodec. Relative to SDCodec, both the encoder and the decoder are replaced with that of DACT.
        
\noindent    \textbf{TUSS-DAC}:
        Proposed cascaded system. We employ the TUSS-Medium model for source separation and DAC as the codec, with the two models trained independently. This system is regarded as the upper bound in terms of reconstruction and separation quality.

\noindent    \textbf{FasTUSS-DACT}:
        Proposed computationally efficient cascaded system.
        We use FasTUSS~\cite{paissan2025fastuss} for source separation and DACT as the codec, with the two models trained independently.

\noindent    \textbf{SUNAC}:
        Proposed integrated system. %
        The encoder architecture is identical to the convolutional part of the encoder in DACT.
        The cross-prompt module consists of a single Transformer layer with 1024 hidden units and 8 attention heads, the conditioning module of a FiLM block with a residual connection, and the target source extraction module of two Transformer layers, each with 1024 hidden units and 8 attention heads.
        Compared with the SDCodec family, SUNAC uses a single RVQ module shared across all domains.
        The decoder is identical to that of DACT.

We use pre-trained models for DAC~\footnote{https://github.com/descriptinc/descript-audio-codec}, TUSS~\footnote{https://github.com/merlresearch/unified-source-separation}, and FasTUSS.
For SDCodec, we used two models: (i) the publicly available pre-trained model\footnote{https://github.com/XiaoyuBIE1994/SDCodec}, trained with source-count probabilities of 0.6, 0.2, and 0.2 for 1, 2, and 3 sources, respectively; and (ii) an SDCodec separately trained with uniform probability of selecting 1, 2, or 3 sources.
The same source-count distribution was also adopted when training the remaining NACs.

For the remaining NACs, including the proposed method, we followed the SDCodec training setup, except that we reduced the batch size to 32 to fit our computational environment.

For SUNAC, we first randomly sample the number of sources $N \in \{1,2,3\}$.
We then select $N$ sources subject to two constraints: the number of \texttt{<Speech>} sources never exceeds two, and \texttt{<Music>} and \texttt{<SFX>} cannot be repeated in the same mixture.

To evaluate performance, we used the SI-SDR and ViSQOL\footnote{https://github.com/google/visqol}
~\cite{chinen2020visqolv3}.
Following~\cite{bie2025learning}, SI-SDR of the separated codec outputs is computed on signals reconstructed by applying the magnitude mask of each separated source to the input mixture.
For ViSQOL, we evaluated the direct output of each decoder.
The evaluation was done on the updated Divide and Remaster (DnR) 
dataset~\cite{petermann2022cocktailforkproblem}, 
where each speech source contains a single speaker, as well as on a similarly derived dataset where the mixtures additionally include an interfering-speaker source for the two-speaker conditions.

\vspace{-.1cm}
\subsection{Experimental results}
\label{subsec:results}
We compare the number of parameters and the computational cost of each method in Table~\ref{table:params_macs}. 
We quantify computational cost by the number of multiply–accumulate operations (MACs). 
Replacing the convolutional components with Transformers reduces the computational cost while keeping the parameter count unchanged (e.g., TUSS~$\rightarrow$~FasTUSS, DAC~$\rightarrow$~DACT, SDCodec~$\rightarrow$~SDCodecT). 
Previous work~\cite{paissan2025fastuss} further showed that, particularly for short audio chunks, most of the compute is spent on convolutions, and that the contributions of the Transformer and convolutional modules remain comparable for 30-s sequences.
Therefore, our proposed SUNAC is more computationally efficient than the conventional SDCodec and the cascaded TUSS–DAC, and it remains more efficient than the lighter cascaded version of FasTUSS–DACT.
All codecs operate at the same bitrate of 6~kbps. However, the SDCodec family employs source-specific codebooks, whereas the other methods share codebooks across sources.

Hereafter, all values are given as mean $\pm$ standard deviation.
The symbol $^\dagger$ denotes a pretrained model.
Table~\ref{tab:reconst_nonmix} reports reconstruction results for isolated sources, 
with comparable performance and no meaningful accuracy gaps across methods.

We evaluated the one-output-per-source-type setting with \texttt{\textless Sp\allowbreak eech\textgreater}, \texttt{<Music>}, \texttt{<SFX>} in Table~\ref{tab:seprate_w/o_perm}. 
The results indicate that the proposed SUNAC achieves performance comparable to SDCodec.
Compared with cascaded approaches, the SDCodec family and SUNAC tend to achieve lower SI-SDR,
which suggests that phase estimation remains challenging. 
However, SUNAC attains VisQOL comparable to FasTUSS-DACT.
Table~\ref{tab:speech_results} reports two-speaker separation results. 
Because the SDCodec family cannot handle multiple sources of the same type,
its performance is substantially lower, 
whereas SUNAC achieves results comparable to the cascaded approaches. 
Table~\ref{tab:two_speech_music_sfx} presents the {\texttt{<Speech>}, \texttt{<Speech>}, \texttt{<Music>}, \texttt{<SFX>}} setting, 
for which TUSS and FasTUSS have been trained but SUNAC has not (we only train SUNAC with up to three prompts). In spite of this handicap, we find SUNAC remains comparable to these cascaded methods.

\vspace{-.2cm}
\section{Conclusion}
We proposed two systems that can generate discrete codes for one or more sources specified by the user from within a mixture: one based on a cascade of separation and NAC, and the other, SUNAC, capable of directly encoding without explicit separation.
In both cases,
a prompt-based extraction module provides flexible control over the number and type of sources.
Experiments show that both systems reliably estimate per-source codes even with multiple sources of the same type, 
a capability unavailable in conventional source-disentangling codecs such as SDCodec. 
Furthermore, SUNAC achieves performance comparable to the cascaded pipeline,
while offering substantially lower computational complexity. 
In future work, we will evaluate the learned disentangled representations on relevant downstream tasks.

\clearpage
\balance
\let\oldthebibliography\thebibliography
\renewcommand{\thebibliography}[1]{%
  \oldthebibliography{#1}%
  \small
  \setlength{\itemsep}{0.75pt}%
  \setlength{\parskip}{0.75pt}%
}

\bibliographystyle{IEEEtran}
\bibliography{codec}

@string{icassp = "Proc. ICASSP"}

@string{interspeech = "Proc. Interspeech"}

@string{iwaenc = "Proc. IWAENC"}

@string{waspaa = "Proc. WASPAA"}

@string{ismir = "Proc. ISMIR"}

@string{ieee-acm-taslp = "IEEE/ACM Trans. Audio, Speech, Lang. Process."}

@string{iclr = "Proc. ICLR"}

@string{neurips = "Proc. NeurIPS"}

@string{aaai = "Proc. AAAI"}

@string{tmlr = "TMLR"}

@inproceedings{kumar2023highfidelity,
      title={High-Fidelity Audio Compression with Improved {RVQGAN}}, 
      author={Rithesh Kumar and Prem Seetharaman and Alejandro Luebs and Ishaan Kumar and Kundan Kumar},
      year={2023},
      Booktitle = neurips,
}

@inproceedings{kong2020hifigan,
  title={{HiFi-GAN}: Generative Adversarial Networks for Efficient and High Fidelity Speech Synthesis}, 
  author={Jungil Kong and Jaehyeon Kim and Jaekyoung Bae},
  year={2020},
  booktitle = neurips,
}

@inproceedings{jang2021univnet,
  title={{UnivNet}: A Neural Vocoder with Multi-Resolution Spectrogram Discriminators for High-Fidelity Waveform Generation}, 
  author={Won Jang and Dan Lim and Jaesam Yoon and Bongwan Kim and Juntae Kim},
  year={2021},
  booktitle = interspeech,
}

@inproceedings{perez2018film,
      title={{FiLM}: Visual Reasoning with a General Conditioning Layer}, 
      author={Ethan Perez and Florian Strub and Harm de Vries and Vincent Dumoulin and Aaron Courville},
      year={2018},
      booktitle=aaai,
      
}

@inproceedings{chinen2020visqolv3,
      title={{ViSQOL} v3: An Open Source Production Ready Objective Speech and Audio Metric}, 
      author={Michael Chinen and Felicia S. C. Lim and Jan Skoglund and Nikita Gureev and Feargus O'Gorman and Andrew Hines},
      year={2020},
      booktitle={Proc. QoMEX}
}

@article{zeghidour2021soundstream,
  title={{SoundStream}: An End-to-End Neural Audio Codec}, 
  author={Zeghidour, Neil and Luebs, Alejandro and Omran, Ahmed and Skoglund, Jan and Tagliasacchi, Marco},
  journal=ieee-acm-taslp,
  volume={30},
  pages={495--507},
  year={2021},
  publisher={IEEE}
}

@article{defossez2022highfi,
    title={High Fidelity Neural Audio Compression},
    author={Alexandre D{\'e}fossez and Jade Copet and Gabriel Synnaeve and Yossi Adi},
    journal=tmlr,
    issn={2835-8856},
    year={2023},
    OPTurl={https://openreview.net/forum?id=ivCd8z8zR2},
}

@article{mousavi2024dasb,
      title={{DASB} - Discrete Audio and Speech Benchmark}, 
      author={Pooneh Mousavi and Luca {Della Libera} and Jarod Duret and Artem Ploujnikov and Cem Subakan and Mirco Ravanelli},
      year={2024},
      eprint={2406.14294},
      archivePrefix={arXiv},
      primaryClass={cs.SD},
      journal={arXiv preprint arXiv:2406.14294},
}

@article{kyutai2024moshi,
      title={Moshi: A speech-text foundation model for real-time dialogue},
      author={Alexandre D\'efossez and Laurent Mazar\'e and Manu Orsini and
      Am\'elie Royer and Patrick P\'erez and Herv\'e J\'egou and Edouard Grave and Neil Zeghidour},
      year={2024},
      eprint={2410.00037},
      archivePrefix={arXiv},
      primaryClass={eess.AS},
      journal={arXiv preprint arXiv:2410.00037},
}

@inproceedings{zhang2024speechtokenizer,
      title={{SpeechTokenizer}: Unified Speech Tokenizer for Speech Large Language Models}, 
      author={Xin Zhang and Dong Zhang and Shimin Li and Yaqian Zhou and Xipeng Qiu},
      year={2024},
      booktitle=iclr,
}

@inproceedings{bie2025learning,
      title={Learning Source Disentanglement in Neural Audio Codec}, 
      author={Xiaoyu Bie and Xubo Liu and Gaël Richard},
      year={2025},
      booktitle=icassp, 
}

@inproceedings{paissan2025fastuss,
      title={{FasTUSS}: Faster Task-Aware Unified Source Separation}, 
      author={Francesco Paissan and Gordon Wichern and Yoshiki Masuyama and Ryo Aihara and François G. Germain and Kohei Saijo and Jonathan {Le Roux}},
      year={2025},
      booktitle=waspaa, 
}

@inproceedings{casebeer2021enhancing,
      title={Enhancing into the codec: Noise Robust Speech Coding with Vector-Quantized Autoencoders}, 
      author={Jonah Casebeer and Vinjai Vale and Umut Isik and Jean-Marc Valin and Ritwik Giri and Arvindh Krishnaswamy},
      year={2021},
      booktitle=icassp, 
}

@inproceedings{yang2021source,
      title={Source-Aware Neural Speech Coding for Noisy Speech Compression}, 
      author={Haici Yang and Kai Zhen and Seungkwon Beack and Minje Kim},
      year={2021},
      booktitle=icassp, 
}

@inproceedings{chae2025towards,
      title={Towards Bitrate-Efficient and Noise-Robust Speech Coding with Variable Bitrate {RVQ}}, 
      author={Yunkee Chae and Kyogu Lee},
      year={2025},
      booktitle=interspeech, 
}

@inproceedings{saijo2025task-aware,
      title={Task-Aware Unified Source Separation}, 
      author={Kohei Saijo and Janek Ebbers and François G. Germain and Gordon Wichern and Jonathan {Le Roux}},
      year={2025},
      booktitle=icassp, 
}

@inproceedings{saijo2024tf-locoformer,
      title={{TF-Locoformer}: Transformer with Local Modeling by Convolution for Speech Separation and Enhancement}, 
      author={Kohei Saijo and Gordon Wichern and François G. Germain and Zexu Pan and Jonathan {Le Roux}},
      year={2024},
      booktitle=iwaenc, 
}

@article{luo2025decodec,
      title={DeCodec: Rethinking Audio Codecs as Universal Disentangled Representation Learners}, 
      author={Xiaoxue Luo and Jinwei Huang and Runyan Yang and Yingying Gao and Junlan Feng and Chao Deng and Shilei Zhang},
      year={2025},
      eprint={2509.09201},
      archivePrefix={arXiv},
      primaryClass={cs.SD},
      journal={arXiv preprint 2509.09201},
}

@inproceedings{pons2024gass,
      title={{GASS}: Generalizing Audio Source Separation with Large-scale Data}, 
      author={Jordi Pons and Xiaoyu Liu and Santiago Pascual and Joan Serrà},
      year={2024},
      booktitle=icassp, 
}

@inproceedings{manilow2020hierarchical,
      title={Hierarchical Musical Instrument Separation}, 
      author={Ethan Manilow and Gordon Wichern and Jonathan {Le Roux}},
      year={2020},
      booktitle=ismir, 
}

@inproceedings{petermann2023hyperbolic,
      title={Hyperbolic Audio Source Separation}, 
      author={Darius Petermann and Gordon Wichern and Aswin Subramanian and Jonathan {Le Roux}},
      year={2023},
      booktitle=icassp 
}

@article{mousavi2025discrete,
      title={Discrete Audio Tokens: More Than a Survey!}, 
      author={Pooneh Mousavi and Gallil Maimon and Adel Moumen and Darius Petermann and Jiatong Shi and Haibin Wu and Haici Yang and Anastasia Kuznetsova and Artem Ploujnikov and Ricard Marxer and Bhuvana Ramabhadran and Benjamin Elizalde and Loren Lugosch and Jinyu Li and Cem Subakan and Phil Woodland and Minje Kim and Hung-yi Lee and Shinji Watanabe and Yossi Adi and Mirco Ravanelli},
      year={2025},
      eprint={2506.10274},
      archivePrefix={arXiv},
      primaryClass={cs.SD},
      journal={arXiv preprint 2506.10274},
}

@inproceedings{leroux2019sdr,
      title={{SDR} - half-baked or well done?}, 
      author={Jonathan {Le Roux} and Scott Wisdom and Hakan Erdogan and John R. Hershey},
      year={2019},
      booktitle=icassp 
}

@Article{Kolbaek2017multitalker,
  author  = {M. Kolb{\ae}k and D. Yu and Z. H. Tan and J. Jensen},
  journal = ieee-acm-taslp,
  title   = {Multitalker Speech Separation With Utterance-Level Permutation Invariant Training of Deep Recurrent Neural Networks},
  year    = {2017},
  month   = jul,
  number  = {10},
  pages   = {1901--1913},
  volume  = {25},
}

@inproceedings{Hershey2016deep,
  author    = {J. R. Hershey and Z. Chen and J. {Le Roux} and S. Watanabe},
  booktitle = icassp,
  title     = {Deep clustering: Discriminative embeddings for segmentation and separation},
  year      = {2016},
}

@inproceedings{petermann2022cocktailforkproblem,
      title={The Cocktail Fork Problem: Three-Stem Audio Separation for Real-World Soundtracks}, 
      author={Darius Petermann and Gordon Wichern and Zhong-Qiu Wang and Jonathan {Le Roux}},
      year={2022},
      booktitle = icassp,
}

@inproceedings{vaswani2023attention,
      title={Attention Is All You Need}, 
      author={Ashish Vaswani and Noam Shazeer and Niki Parmar and Jakob Uszkoreit and Llion Jones and Aidan N. Gomez and Lukasz Kaiser and Illia Polosukhin},
      year={2023},
      booktitle = neurips,
}

@inproceedings{hu2025salmduplex,
      title={{SALM-Duplex}: Efficient and Direct Duplex Modeling for Speech-to-Speech Language Model}, 
      author={Ke Hu and Ehsan Hosseini-Asl and Chen Chen and Edresson Casanova and Subhankar Ghosh and Piotr Żelasko and Zhehuai Chen and Jason Li and Jagadeesh Balam and Boris Ginsburg},
      year={2025},
      booktitle = interspeech,
}

@inproceedings{shang2024endtoendspeechsummarization,
      title={An End-to-End Speech Summarization Using Large Language Model}, 
      author={Hengchao Shang and Zongyao Li and Jiaxin Guo and Shaojun Li and Zhiqiang Rao and Yuanchang Luo and Daimeng Wei and Hao Yang},
      year={2024},
      booktitle = interspeech,
}

@inproceedings{matsuura2024sentencewises,
      title={Sentence-wise Speech Summarization: Task, Datasets, and End-to-End Modeling with LM Knowledge Distillation}, 
      author={Kohei Matsuura and Takanori Ashihara and Takafumi Moriya and Masato Mimura and Takatomo Kano and Atsunori Ogawa and Marc Delcroix},
      year={2024},
      booktitle = interspeech,
}

@inproceedings{wang2024leveraginglanguagemodel,
      title={Leveraging Language Model Capabilities for Sound Event Detection}, 
      author={Hualei Wang and Jianguo Mao and Zhifang Guo and Jiarui Wan and Hong Liu and Xiangdong Wang},
      year={2024},
      booktitle = interspeech,
}

@INPROCEEDINGS{yin2025exploring,
  author={Yin, Han and Bai, Jisheng and Xiao, Yang and Wang, Hui and Zheng, Siqi and Chen, Yafeng and Das, Rohan Kumar and Deng, Chong and Chen, Jianfeng},
  title={Exploring Text-Queried Sound Event Detection with Audio Source Separation}, 
  year={2025},
  booktitle = icassp,
}

@inproceedings{zhang2025instructmusicgen,
      title={{Instruct-MusicGen}: Unlocking Text-to-Music Editing for Music Language Models via Instruction Tuning},
      author={Yixiao Zhang and Yukara Ikemiya and Woosung Choi and Naoki Murata and Marco A. Martínez-Ramírez and Liwei Lin and Gus Xia and Wei-Hsiang Liao and Yuki Mitsufuji and Simon Dixon},
      year={2025},
      booktitle = ismir,
}

@inproceedings{wang2025usam,
      title={{U-SAM}: An audio language Model for Unified Speech, Audio, and Music Understanding}, 
      author={Ziqian Wang and Xianjun Xia and Xinfa Zhu and Lei Xie},
      year={2025},
      booktitle = interspeech,
}

@inproceedings{ohashi2025jmoshi,
      title={Towards a {J}apanese Full-duplex Spoken Dialogue System}, 
      author={Atsumoto Ohashi and Shinya Iizuka and Jingjing Jiang and Ryuichiro Higashinaka},
      year={2025},
      booktitle = interspeech,
}

@article{su2023roformer,
      title={RoFormer: Enhanced Transformer with Rotary Position Embedding}, 
      author={Jianlin Su and Yu Lu and Shengfeng Pan and Ahmed Murtadha and Bo Wen and Yunfeng Liu},
      year={2023},
      eprint={2104.09864},
      archivePrefix={arXiv},
      primaryClass={cs.CL},
      journal={arXiv preprint 2104.09864},
}

\end{document}